\documentclass[article]{elsart}
\usepackage{amssymb}
\usepackage{epsfig}
\usepackage{units}
\begin{document}

\begin{frontmatter}

\title{ Determination of quadrupole strengths in the $\gamma^*p\rightarrow
\Delta(1232)$ transition at $Q^2= 0.20$ (GeV/c)$^2$}

\author[athens]{N.F.~Sparveris},
\author[mainz]{P.~Achenbach},
\author[mainz]{C.~Ayerbe Gayoso},
\author[mainz]{D.~Baumann},
\author[mainz]{J.~Bernauer},
\author[mit]{A. M. Bernstein},
\author[mainz]{R.~B\"ohm},
\author[zagreb]{D.~Bosnar},
\author[mit]{T. Botto},
\author[athens]{A.~Christopoulou},
\author[uk]{D.~Dale},
\author[mainz]{M.~Ding},
\author[mainz]{M.~O.~Distler},
\author[mainz]{L.~Doria},
\author[mainz]{J.~Friedrich},
\author[athens]{A.~Karabarbounis},
\author[zagreb]{M.~Makek},
\author[mainz]{H.~Merkel},
\author[mainz]{U.~M\"uller},
\author[riken]{I.~Nakagawa},
\author[mainz]{R.~Neuhausen},
\author[mainz]{L.~Nungesser},
\author[athens]{C.N.~Papanicolas\corauthref{cor}},
        \corauth[cor]{Corresponding author.}
        \ead{cnp@iasa.gr}
\author[mainz]{A.~Piegsa},
\author[mainz]{J.~Pochodzalla},
\author[ulj]{M.~Potokar},
\author[mainz]{M.~Seimetz},
\author[ulj]{S.~\v Sirca},
\author[mit]{S.~Stave},
\author[athens]{S.~Stiliaris},
\author[mainz]{Th.~Walcher}
and
\author[mainz]{M.~Weis}

\address[athens]{Institute of Accelerating Systems and Applications
and Department of Physics, University of Athens, Athens, Greece}

\address[mainz]{Institut f\"ur Kernphysik, Universit\"at Mainz,
  Mainz, Germany}

\address[riken]{Radiation Laboratory, RIKEN, 2-1
Hirosawa, Wako, Saitama 351-0198, Japan}

\address[mit]{Department of Physics, Laboratory for Nuclear Science
and Bates Linear Accelerator Center, Massachusetts Institute of
Technology, Cambridge, Massachusetts 02139, USA}

\address[zagreb]{Department of Physics, University of Zagreb, Croatia}

\address[uk]{Department of Physics and Astronomy, University of
  Kentucky, Lexington, Kentucky 40206 USA}

\address[ulj]{Institute Jo\v zef Stefan, University of Ljubljana,
  Ljubljana, Slovenia}

\date{\today}

\begin{abstract}

We report new precise p$(\vec{e},e^\prime p)\pi^0$ measurements at
the peak of the $\Delta^{+}(1232)$ resonance at
$Q^2=\unit[0.20](GeV/c)^2$ performed at the Mainz Microtron (MAMI).
The new data are sensitive to both the electric quadrupole ($E2$)
and the coulomb quadrupole ($C2$) amplitudes of the $\gamma^*
N\rightarrow\Delta$ transition. They yield precise quadrupole to
dipole amplitude ratios CMR $= (-5.09 \pm 0.28_{stat+sys}\pm
0.30_{model})\%$ and EMR $= (-1.96 \pm 0.68_{stat+sys} \pm
0.41_{model})\%$ for $M^{3/2}_{1+} = (39.57 \pm 0.75_{stat+sys}\pm
0.40_{model})(10^{-3}/m_{\pi^+})$. The new results are in
disagreement with Constituent Quark Model predictions and in
qualitative agreement with models that account for mesonic
contributions, including recent Lattice calculations. They thus give
further credence to the conjecture of deformation in hadronic
systems favoring the attribution of the origin of deformation to the
dominance of mesonic effects.

\end{abstract}

\end{frontmatter}


In recent years an extensive effort has been focused on identifying
and understanding the origin of possible non-spherical components in
the nucleon wavefunction
\cite{Ru75,is82,pho2,pho1,frol,pos01,merve,bart,Buuren,joo,spaprc,kun00,spaprl,stave,dina,sato,dmt00,kama,mai00,multi,said}.
The complex quark-gluon and meson cloud dynamics of hadrons give
rise to non-spherical components in their wavefunction which in a
classical limit and at large wavelengths will correspond to a
"deformation". The most direct evidence of deformation is provided
through the measurement of the spectroscopic quadrupole moment which
for the proton, the only stable hadron, vanishes identically because
of its spin 1/2 nature. Instead the signature of the deformation of
the proton is sought in the presence of resonant quadrupole
amplitudes $(E^{3/2}_{1+}, S^{3/2}_{1+})$ in the predominantly
magnetic dipole ($M^{3/2}_{1+}$) $\gamma^* N\rightarrow \Delta$
transition \cite{rev2,rev3}. Non vanishing resonant quadrupole
amplitudes will signify that either the proton or the
$\Delta^{+}(1232)$ or more likely both are deformed. The ratios CMR
$= Re(S^{3/2}_{1+}/M^{3/2}_{1+})$ and EMR $=
Re(E^{3/2}_{1+}/M^{3/2}_{1+})$ are routinely
\cite{dmt00,kama,mai00,multi} used to present the relative magnitude
of the amplitudes of interest.

Depending on the interpretative framework adopted the origin of the
deformation is attributed to a number of different processes. In the
constituent-quark picture of hadrons, it arises as a consequence of
the non-central color-hyperfine interaction among quarks
\cite{Ru75,is82}, while in dynamical models of the $\pi N$ system,
deformation also arises from the asymmetric coupling of the pion
cloud to the quark core. Our current understanding of the nucleon
suggests that at long distances (low momenta) the pionic cloud
effect dominates while at short distances (high momenta) intra-quark
forces predominate. Recent precise experimental results
\cite{pho2,pho1,frol,pos01,merve,bart,Buuren,joo,spaprc,kun00,spaprl,stave}
are in reasonable agreement with predictions of models invoking
deformation while at the same time exclude all nucleon models that
assume sphericity for the proton and the delta, and thus they
confirm the deviation from spherical shape. With the existence of
non-spherical components in the nucleon wavefunction well
established, recent investigations have focused on understanding the
various mechanisms that could generate it.

The measurements reported in this paper have been performed in the
low momentum transfer region at $Q^2=\unit[0.20](GeV/c)^2$ where the
pionic contribution is expected to dominate \cite{sato,dmt00,kama}.
The new data taken together with the measurements at the photon
point \cite{pho2,pho1}, at $Q^2= 0.06$ (GeV/c)$^2$ \cite{sean} and
$Q^2=\unit[0.127](GeV/c)^2$ \cite{pos01,spaprl} explore the low
$Q^2$ dependence of the quadrupole amplitudes  providing valuable
insights into the mechanism that generates the deformation.

The cross section of the $p(\vec{e},e^\prime p)\pi^0$ reaction is
sensitive to five independent partial responses
($\sigma_{T},\sigma_{L},\sigma_{LT}, \sigma_{TT}$ and
$\sigma_{LT'}$) \cite{multi} :

\begin{eqnarray}
 \frac{d^5\sigma}{d\omega d\Omega_e d\Omega^{cm}_{pq}} & = & \Gamma (\sigma_{T} + \epsilon{\cdot}\sigma_L
  - v_{LT}{\cdot}\sigma_{LT}{\cdot}\cos{\phi_{pq}^{*}} \nonumber \\
 & &   +\epsilon{\cdot}\sigma_{TT}{\cdot}\cos{2\phi_{pq}^{*}} - h {\cdot} p_e {\cdot} v_{LT'}{\cdot}\sigma_{LT'}{\cdot}\sin{\phi_{pq}^{*}}) \nonumber
\label{equ:cros}
\end{eqnarray}

where the kinematic factors $v_{LT}=\sqrt{2\epsilon(1+\epsilon)}$
and $v_{LT'}=\sqrt{2\epsilon(1-\epsilon)}$, $\epsilon$ is the
transverse polarization of the virtual photon, $\Gamma$ the virtual
photon flux, $h=\pm1$ is the electron helicity, $p_e$ is the
magnitude of the longitudinal electron polarization, and
$\phi_{pq}^{*}$ is the proton azimuthal angle with respect to the
electron scattering plane. The virtual photon differential cross
sections ($\sigma_{T},\sigma_{L},\sigma_{LT}, \sigma_{TT}$ and
$\sigma_{LT'}$) are all functions of the center of mass energy W,
the four momentum transfer squared $Q^2$, and the proton center of
mass polar angle $\theta_{pq}^{*}$ (measured from the momentum
transfer direction) \cite{multi}.

The $\sigma_{0}=\sigma_{T}$+$\epsilon\cdot\sigma_{L}$ response is
dominated by the $M_{1+}$ resonant multipole. The interference of
the $C2$ and $E2$ amplitudes with the $M1$ dominates the
Longitudinal~-~Transverse (LT) and Transverse~-~Transverse (TT)
responses respectively.  The $\sigma_{LT'}$ response \cite{multi}
provides sensitivity to background contributions \cite{mande}
primarily through the $Im(S^*_{0+}M_{1+})$ term; as a result the
real part of $S_{0+}$ is manifested through interference with the
dominant imaginary part of $M_{1+}$.

The experiment was performed at the Mainz Microtron using the A1
magnetic spectrometers \cite{spectr} in an arrangement identical to
that reported in \cite{stave,sean}. An $\unit[855] MeV$ polarized
electron beam  was employed on a liquid-hydrogen target. Beam
polarization was measured periodically with a M$\o$ller polarimeter
to be $\approx 75\%$. The beam average current was $\unit[25] \mu
A$. Electrons and protons were detected in coincidence with
spectrometers A and B respectively. Both spectrometers use two pairs
of vertical drift chambers respectively for track reconstruction and
two layers of scintillator detectors for timing information and
particle identification \cite{spectr}. Spectrometer B allows
out-of-plane detection capability of up to $10^\circ$ with respect
to the horizontal plane which allows access to the fifth response.
Measurements with the proton spectrometer at three different
azimuthal angles, $\phi_{pq}^{*}$, for the same central kinematics
in $W, Q^2$ and $\theta_{pq}^{*}$ allowed the extraction of all
three unpolarized partial cross sections $\sigma_{TT}$,
$\sigma_{LT}$ and $\sigma_{0}=\sigma_{T}+\epsilon \cdot \sigma_{L}$.

Measurements were performed at central kinematics of
$W=\unit[1221]MeV$ and $Q^2=\unit[0.20](GeV/c)^2$ and at the proton
angles of $\theta_{pq}^{*}=0^\circ$, 33$^\circ$ and $57^\circ$. At
each $\theta_{pq}^{*}= 33^\circ$ and $57^\circ$ kinematics, the
proton spectrometer was sequentially placed at three different
azimuthal angles $\phi_{pq}^{*}$. At the $\theta_{pq}^{*}= 33^\circ$
setup, the proton spectrometer was placed at $\phi_{pq}^{*}=
0^\circ$, 90$^\circ$ and $180^\circ$ while at $\theta_{pq}^{*}=
57^\circ$ due to space limitations caused by the beam line it was
placed at $\phi_{pq}^{*}= 38^\circ$, 142$^\circ$ and $180^\circ$.
Thus in both cases it became possible to isolate $\sigma_{TT}$,
$\sigma_{LT}$ and $\sigma_{0}=\sigma_{T}+ \epsilon \cdot \sigma_{L}$
partial cross sections. The out of plane measurements allowed us to
extract the $\sigma_{LT'}$ cross section at both $\theta_{pq}^{*}=
33^\circ$ and $57^\circ$ kinematics. In the case of the
$\theta_{pq}^{*}= 33^\circ$ kinematics the extensive phase space
coverage of the spectrometers as well as the sufficient statistics
allowed the extraction of the partial cross sections at the proton
angles of $\theta_{pq}^{*}= 27^\circ$ and $40^\circ$ as well.

Systematic uncertainties were reduced by using Spectrometer C
throughout the experiment as a luminosity monitor. Elastic
scattering data from H and $^{12}C$ for calibration purposes were
taken at 600 MeV. Cross sections were measured at $Q^2$ = 0.127
$(GeV/c)^2$ \cite{sean} at kinematics similar to those of Bates
\cite{spaprl}; the two sets of data are statistically compatible
insuring that data obtained from the two laboratories can be
meaningfully compared as done in Fig.~2.

In Fig.~1 we present the experimental results for $\sigma_o =
\sigma_{T}$+$\epsilon{\cdot}\sigma_{L}$, $\sigma_{LT}$,
$\sigma_{TT}$ and $\sigma_{LT'}$ compared with the SAID multipole
analysis \cite{said}, the phenomenological model MAID 2003
\cite{mai00,kama}, the dynamical model calculations of Sato-Lee
\cite{sato} and of DMT (Dubna - Mainz - Taipei) \cite{dmt00} and the
ChEFT calculation of Pascalutsa and Vanderhaegen \cite{pv}. The SAID
multipole analysis \cite{said} describes the data adequately with an
overall tendency to overestimate them; it fails to describe the
$\sigma_{0}$ measurement at $\theta_{pq}^{*}= 0^\circ$. The MAID
model \cite{kama,mai00} which offers a flexible phenomenology and
which was the most successful in describing the
$Q^2=\unit[0.127](GeV/c)^2$ data \cite{spaprl} overestimates
$\sigma_o$, $\sigma_{LT}$ and $\sigma_{LT'}$ thus indicating that a
re-adjustment is needed both in the resonant  and in the background
amplitudes. The Sato-Lee \cite{sato} and DMT \cite{dmt00} dynamical
models provide a nucleon description which explicitly incorporates
the physics of the pionic cloud. Both calculate the resonant
channels from dynamical equations. DMT uses the background
amplitudes of MAID with some small modifications. Sato-Lee calculate
all amplitudes consistently within the same framework with only
three free parameters. Both find that a large fraction of the
quadrupole multipole strength arises due to the pionic cloud with
the effect reaching a maximum value in the region $Q^2=\unit[0.15]
(GeV/c)^2$. Sato-Lee offers a good description of the data slightly
overestimating $\sigma_o$ for proton angles above $\theta_{pq}^{*}=
50^\circ$. DMT disagrees with the $\sigma_{LT}$ measurements
indicating that the Coulomb quadrupole amplitude is overestimated.
The chiral perturbation calculation of Pascalutsa and Vanderhaegen
\cite{pv} is also presented in Fig.~1. In addition to the predicted
central value an estimate of the model uncertainty is provided by
calculating the magnitude of the next order terms in the chiral
expansion. Clearly the rather large uncertainties that are
associated with this calculation render the prediction compatible
with the measurements while at the same time making the need for the
next order calculation obvious. It would be very useful if all of
the model calculations would also provide similar model errors which
will render the comparison of theory and experiment far more
meaningful.

The resonant $M_{1+}^{3/2}$, $S_{1+}^{3/2}$ and $E_{1+}^{3/2}$ have
been extracted from the measured partial cross sections. A Truncated
Multipole Expansion (TME) fit, in which all background contributions
are set equal to zero, has been performed fitting all three resonant
amplitudes of interest while results have been extrapolated to
$W=\unit[1232]MeV$. The resulting TME fit is characterized by high
$\chi^2$ (see Table~\ref{tab:fit}) demonstrating the inadequacy of
this approach due to the neglecting of the important background
amplitude contributions. A fit of the three resonant amplitudes has
also been performed while taking into account the contributions of
background amplitudes from MAID, DMT, SAID and Sato-Lee model
predictions; reasonable $\chi^2$ have been obtained. All
3-parameter-fit results are presented in Table~\ref{tab:fit} in
order to exhibit the effect of the model dependence on the
extraction of the amplitudes. The adopted values for the extracted
multipoles (and their ratios) shown in Table~\ref{tab:resu}, result
from the wighted average of the four model fits of
Table~\ref{tab:fit}. We adopt the RMS deviation of the fitted
central values as indicative of the model uncertainty. The extracted
value of $(-5.09 \pm 0.28_{stat+sys})\%$ for the CMR is found to be
in good agreement with the value of $(-5.45 \pm 0.42_{stat+sys})\%$
derived from a MAMI asymmetry $p(e,e^{'}p)\pi^0$ measurement
\cite{elsner} where the result is extracted from a MAID re-fit to
the data. EMR is precisely determined for the first time at this
$Q^2$. A precise measurement of the $M_{1+}^{3/2}$ is also provided
through this experiment.

Table~\ref{tab:resu} presents the derived values along with the
respective values of the various model predictions; it is evident
that for the resonant amplitudes an overall consistency in terms of
sign and magnitude among the models considered has emerged.
Nevertheless further refinements are needed in order to describe the
data accurately and to provide a consistent value for both CMR and
EMR. SAID overestimates the data and in particular $\sigma_{0}$ and
it underestimates EMR most likely due to the influence of older data
in its data base. At this $Q^2$ the Sato-Lee model provides the best
description of the experimentally measured responses; it is no
surprise that its values for the CMR and EMR are very close to the
experimentally derived ones.

The results reported here, taken together with the results from the
photon point and from $Q^2=\unit[0.06] (GeV/c)^2$ \cite{stave} and
$Q^2=\unit[0.127](GeV/c)^2$ \cite{spaprl}, complete the experimental
investigation of the issue of nucleon deformation at low $Q^2$. They
allow us to draw conclusions on the mechanism behind the deviation
from sphericity, in particular on the role of the pion cloud a
manifestly low $Q^2$ mechanism. The derived $M_{1+}$, EMR and CMR
values vary smoothly (see Fig.~2) as expected. The SAID, MAID, DMT
and Sato-Lee models are in qualitative agreement with the
experimental results; detailed improvements could and should be
implemented as commented above and in previous publications. The
results from constituent quark models are known to considerably
deviate from the experimental results failing to describe the
resonant quadrupole amplitudes and to substantially underestimate
the dominant $M_{1+}$. Two representative calculations are shown in
Fig.~2, that of Capstick \cite{is82} and of the hypercentral quark
model (HQM) \cite{hqm}, which fail to describe the data. It
demonstrates that the color hyperfine interaction is inadequate to
explain the effect at least at large distances. The recent effective
field theoretical (chiral) calculations \cite{pv,hemmert}, that are
solidly based on QCD, also successfully account for the magnitude of
the effects giving further credence to the dominance of the meson
cloud effect, although as commented above the next order calculation
is required.

Finally, recent results from lattice QCD \cite{dina} are also of
special interest since they are for the first time accurate enough
to allow a comparison to experiment. The chirally extrapolated
\cite{pv} values of CMR and EMR are found to be non zero and
negative in the low $Q^2$ region, in qualitative agreement with the
experimental results, thus linking the experimental evidence for
deformation directly to QCD. This impressive success highlights the
importance of pursuing this avenue with urgency with more precise
lattice results using lighter quark masses and further refining the
chiral extrapolation procedure.

In conclusion, the data presented here provide a precise
determination of both CMR and EMR at $Q^2=\unit[0.20](GeV/c)^2$
while the $M^{3/2}_{1+}$ amplitude is simultaneously being
determined. The non zero values of the resonant quadrupole
amplitudes determined support the conjecture of nucleon deformation.
The new data along with those at $Q^2=\unit[0.127] (GeV/c)^2$
\cite{pos01,spaprl} and the ones at $Q^2=\unit[0.06](GeV/c)^2$
\cite{stave} and the photon point provide a clear and consistent
picture of the quadrupole amplitudes at low $Q^2$; they nicely match
the JLab data at intermediate $Q^2$ \cite{joo}. The quadrupole to
dipole amplitude ratios are found to be bigger by an order of
magnitude than the values predicted by quark models on account of
the noncentral color-hyperfine interaction \cite{is82} and
consistent in magnitude with the ones predicted from models that
take into account the mesonic degrees of freedom
\cite{sato,dmt00,kama,pv,hemmert} reinforcing similar conclusions
reached at $Q^2=\unit[0.127](GeV/c)^2$ and
$Q^2=\unit[0.06](GeV/c)^2$. Recent chiral effective
\cite{pv,hemmert} and lattice calculations with a chiral
extrapolation to the physical pion mass  \cite{dina} which provide a
direct link to QCD are in agreement with the experimental values.

We would like to thank the MAMI accelerator group and the MAMI
polarized beam group for the excellent beam quality combined with a
continuous high polarization. We would also like to thank Drs
L.~Tiator, S.~Kamalov, S.~Yang, T.-S.H.~Lee, M.~Vanderhaegen and
V.~Pascalutsa for their valuable suggestions and calculations. This
work is supported at Mainz by the Sonderforschungsbereich 443 of the
Deutsche Forschungsgemeinschaft (DFG) and by the program PYTHAGORAS
co-funded by the European Social Fund and National Resources (EPEAEK
II).


\newpage

\begin{figure}[p]
\begin{center}
\epsfig{file=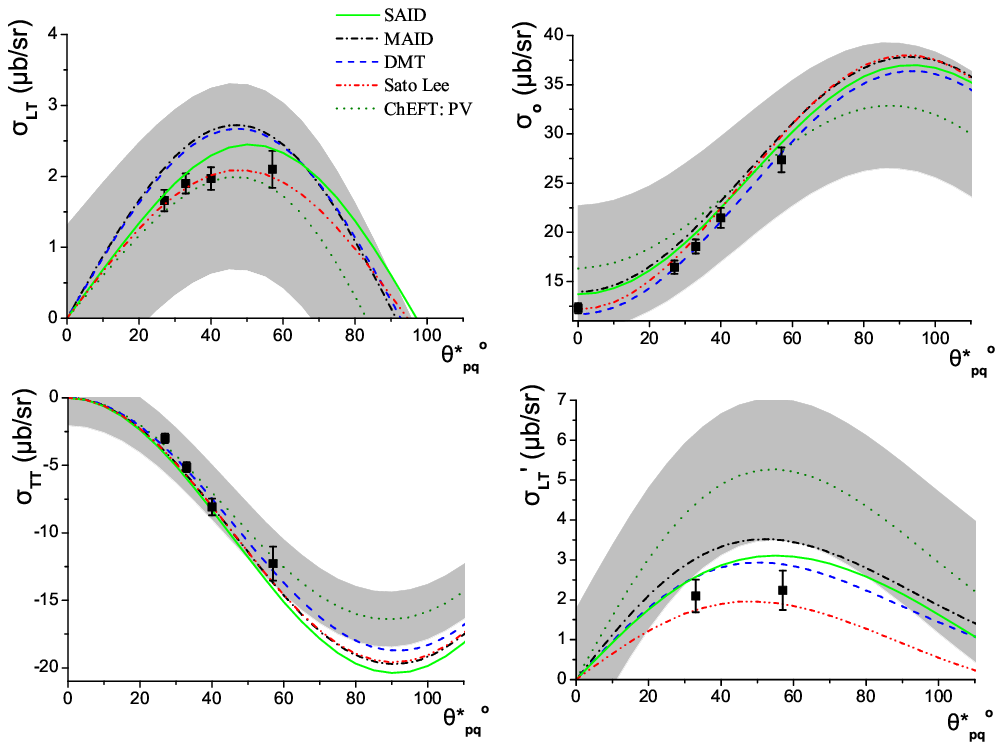,width=17.0cm}
\end{center}
\caption{The measured $\sigma_{0} = \sigma_{T} + \epsilon \cdot
\sigma_{L}$, $\sigma_{LT}$, $\sigma_{TT}$ and $\sigma_{LT'}$ partial
cross sections as a function of $\theta^*_{pq}$ at central
kinematics of $W=\unit[1221]MeV$ and $Q^2=\unit[0.20](GeV/c)^2$. The
theoretical predictions of MAID, DMT, SAID, Sato-Lee and the ChEFT
of Pascalutsa and Vanderhaegen (with the corresponding error band)
are also presented.}
\end{figure}

\begin{figure}[p]
\begin{center}
\epsfig{file=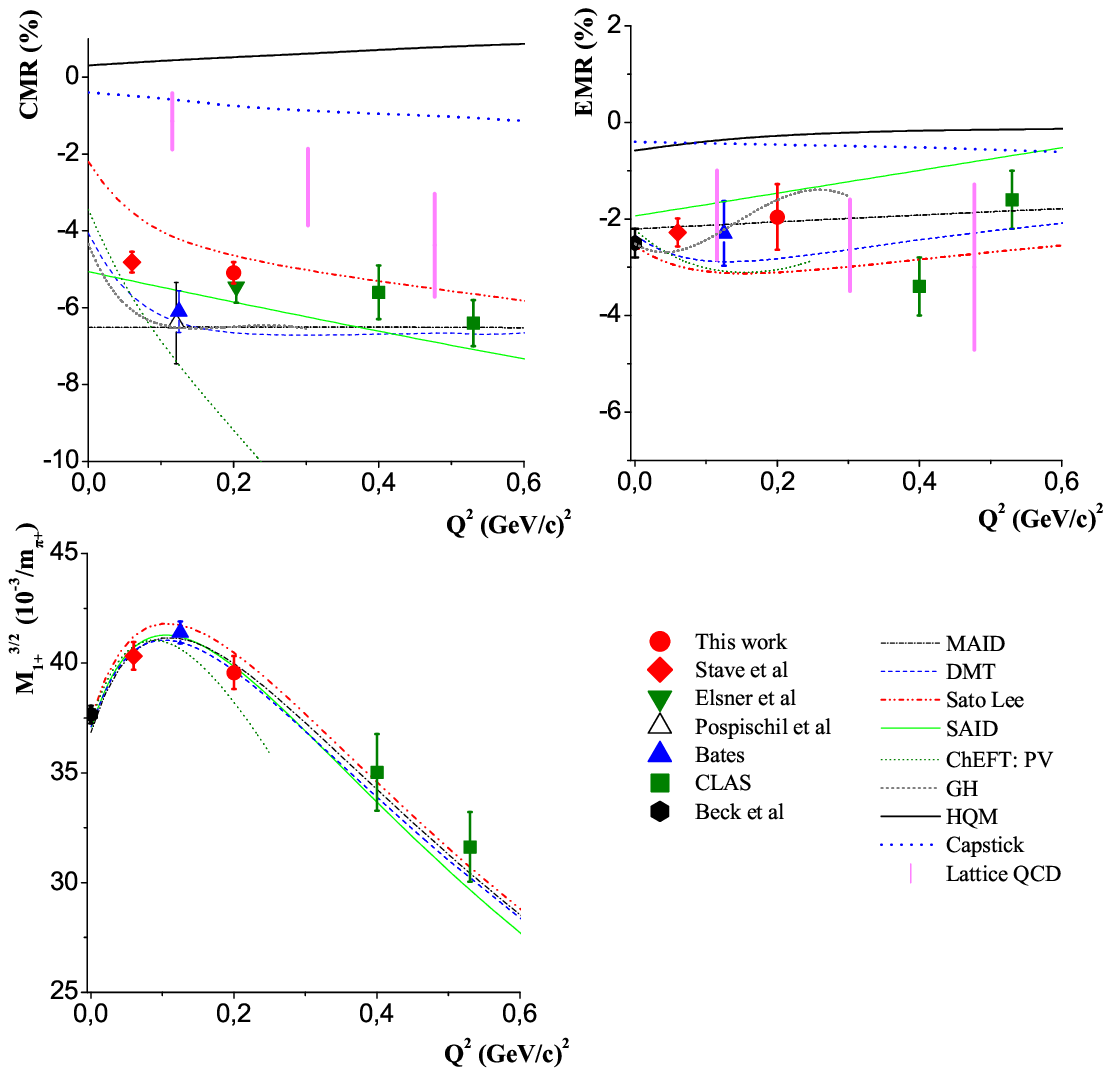,width=17.0cm}
\end{center}
\caption{The extracted values for CMR, EMR and $M_{1+}^{3/2}$
(errors include statistical and systematic uncertainties) as a
function of $Q^2$. The theoretical predictions of MAID, DMT, SAID,
Sato-Lee, ChEFT of Pascalutsa-Vanderhaegen and the Gail-Hemmert and
the experimental results from
\cite{pho1,pos01,joo,spaprl,stave,elsner} are shown.}
\end{figure}

\begin{center}
\begin{table}
\caption{Values of CMR, EMR and $M_{1+}^{3/2}$ at $W=\unit[1232]MeV$
and $Q^2=\unit[0.20](GeV/c)^2$ extracted from the experimental data
performing I) a  Truncated Multipole Expansion (TME) fit to the data
and II) a 3-parameter-fit (resonant amplitudes only) using the
background contributions from the MAID, DMT, SAID and Sato-Lee model
predictions respectively.}
\begin{tabular}{|c|cccc|}
\hline
                    &  CMR $(\%)$          & EMR $(\%)$        & $M_{1+}^{3/2} (10^{-3}/m_{\pi^+})$  & $(\chi^2/d.o.f.)$ \\
\hline
TME                 &  $-5.62 \pm 0.26$    & $-3.00 \pm 0.67$  &  $39.51 \pm 0.75 $ & 4.75 \\
\hline
$fit_{(MAID)}$   &  $-5.50 \pm 0.29$    & $-2.36 \pm 0.69$  &  $39.43 \pm 0.75 $ & 1.31 \\
$fit_{(DMT)} $   &  $-5.04 \pm 0.27$    & $-1.75 \pm 0.67$  &  $39.84 \pm 0.75 $ & 0.82 \\
$fit_{(SAID)}$   &  $-4.68 \pm 0.28$    & $-1.41 \pm 0.67$  &  $38.89 \pm 0.76 $ & 1.52 \\
$fit_{(SL)}  $   &  $-5.11 \pm 0.27$    & $-2.24 \pm 0.69$  & $39.76
\pm 0.75 $ &  0.89 \\ \hline
\end{tabular}
\label{tab:fit}
\end{table}
\end{center}

\begin{center}
\begin{table}
\caption{The experimentally derived values of CMR, EMR and
$M_{1+}^{3/2}$ at $W=\unit[1232]MeV$ and $Q^2=\unit[0.20](GeV/c)^2$
are presented (as extracted from the average of the 3-parameter-fits
to the data (listed in Table~\ref{tab:fit}) which include the
background amplitude contributions from the MAID, DMT, SAID and SL
models). The first errors correspond to the experimental
statistical+systematic fitting uncertainties while the second one
corresponds to the model uncertainties. The initial model values for
the MAID, DMT, SAID, Sato-Lee, Pascalutsa-Vanderhaeghen ChEFT and
the Gail-Hemmert predictions for the amplitudes of interest are also
presented in the table for comparison.}
\begin{tabular}{|c|ccc|}
\hline
                    &  CMR $(\%)$          & EMR $(\%)$        & $M_{1+}^{3/2} (10^{-3}/m_{\pi^+})$ \\
\hline
$ experiment $      &  $-5.09 \pm 0.28 \pm 0.30 $    & $-1.96 \pm 0.68 \pm 0.41 $  &  $39.57 \pm 0.75 \pm 0.40 $ \\
MAID                &  $-6.50                   $    & $-2.06                   $  &  $39.98          $ \\
DMT                 &  $-6.68                   $    & $-2.88                   $  &  $39.85          $ \\
SAID                &  $-5.01                   $    & $-0.60                   $  &  $39.45          $ \\
Sato-Lee            &  $-4.58                   $    & $-3.11                   $  &  $40.48          $ \\
ChEFT:PV            &  $-9.19 \pm 3.00          $    & $-3.05 \pm 1.20          $  &  $38.22 \pm 5.10 $ \\
GH                  &  $-6.50                   $    & $-1.60                   $  &  $               $ \\
\hline
\end{tabular}
\label{tab:resu}
\end{table}
\end{center}


\end{document}